\documentclass[12pt]{article}
\usepackage{latexsym}
\usepackage{axodraw}
\usepackage{epsfig}
\usepackage{amsfonts}
\usepackage{array,longtable}
\begin{document}

 \title{\bf Renormalizable Expansion for Nonrenormalizable Theories:\\
 II. Gauge Higher Dimensional Theories}
\author{D.I.Kazakov}
\date{}
\maketitle \vspace{-0.8cm}
\begin{center}
Bogoliubov Laboratory of Theoretical Physics, Joint
Institute for Nuclear Research, Dubna, Russia \\
Institute for Theoretical and Experimental Physics, Moscow, Russia\\
e-mail: KazakovD@theor.jinr.ru \\[0.5cm]

{\large and G.S.Vartanov}

\vspace{0.8cm}
Museo Storico della Fisica e Centro Studi e Ricerche "Enrico Fermi",
Rome, Italy\\
Bogoliubov Laboratory of Theoretical Physics, Joint
Institute for Nuclear Research, Dubna, Russia \\[0.1cm]
e-mail: Vartanov@theor.jinr.ru\\[0.1cm]
\end{center}

\begin{abstract}
The previously developed renormalizable perturbative 1/N-expansion
in higher dimensional scalar field theories is extended to gauge
theories with fermions. It is based on the $1/N_f$-expansion and
results in a logarithmically divergent perturbation theory in
arbitrary high odd space-time dimension. Due to the
self-interaction of non-Abelian fields the proposed recipe
requires some modification which, however, does not change the
main results. The new effective coupling is dimensionless and is
running in accordance with the usual RG equations. The
corresponding beta function is calculated in the leading order and
is nonpolynomial in effective coupling. The original dimensionful
gauge coupling plays a role of mass and is also logarithmically
renormalized. Comments on the unitarity of the resulting theory
are given.
\end{abstract}

\section{Introduction}

In our previous paper~\cite{KVI} (hereafter [I]) we constructed
renormalizable perturbative 1/N expansion in scalar field theories
in higher dimensions. It was demonstrated that the 1/N
expansion\footnote{for review of the $1/N$ expansion
see~\cite{Justin,Hooft0,Makeenko}} contrary to the usual
perturbation theory is renormalizable~\cite{Parisi,Schnitzer},
leads to a finite number of logarithmically divergent terms and
can be made finite by renormalization of the fields and a new
dimensionless coupling. The original dimensionful coupling does
not serve as an expansion parameter anymore and plays the role of
mass which is also logarithmically divergent and multiplicatively
renormalized. Within the dimensional regularization
technique~\cite{reg} we performed the renormalization procedure in
scalar theories in arbitrary odd space-time dimension and
calculated a few terms of the 1/N expansion. Even dimensions, in
principle, can also be treated by this method; however, they lead
to some complications due to the appearance of log terms. We did
not consider them in [I] and do not do it here.

The main recipe of the 1/N expansion is the following (see for
example~\cite{Arefeva}): one considers a theory with an
N-component field interacting with some real or auxiliary singlet
field and calculates the propagator of this singlet field. If one
divides the coupling of their interaction by $\sqrt{N}$, then any
simple virtual loop in this propagator is suppressed by a factor
of 1/N which is cancelled by an opposite factor coming from the
closed loop of the N-component field. Thus, each of this bubbles
has no 1/N suppression and in the leading order one has to sum
them up. The resulting "zeroth" order propagator contains the
polarization operator, which is an increasing function of momenta,
in the denominator. This new propagator has to be used now in all
Feynman diagrams order by order in 1/N. Since it decreases faster
than the usual one, it improves the convergence of  the integrals
and leads to logarithmically divergent diagrams even in arbitrary
high extra dimensions.

The aim of this paper is to demonstrate how the advocated
procedure works in gauge theories which contain some peculiarities
due to the gauge invariance. We show that indeed in the case of an
Abelian theory the procedure is straightforward while in the
non-Abelian case it needs some modification due to the presence of
the triple and quartic interactions of gauge fields.

When summing up the vacuum polarization diagrams to the denominator, one faces the
problem of unitarity due to the presence of the imaginary part in the polarization
operator and possible poles in the complex plane. This problem is common to any
realization of the 1/N expansion and deserves a special
treatment~\cite{Schnitzer,SchnitzerA}. We briefly comment on this problem below.

\section{QED}


Let us start with the usual QED with $N_f$ fermion fields in $D$ dimensions, where $D$
takes an arbitrary odd value. The Lagrangian looks like
\begin{eqnarray}\label{l}
  {\cal L} &=&
    - \frac 14 (\partial_\mu A_{\nu} - \partial_\nu A_{\mu})^2 -
    \frac {1}{2\alpha} (\partial_{\mu}A_{\mu})^2 + i \bar{\psi_i} \hat{\partial}
  \psi_i - m \bar{\psi_i} \psi_i
  + \frac{e}{\sqrt{N_f}} \bar{\psi_i} \hat{A} \psi_i.
\end{eqnarray}

According to the general strategy, we now have to consider the
photon propagator. Since due to the gauge invariance the
polarization operator is transverse, it is useful to consider a
transverse (Landau) gauge. This is not necessary but simplifies
the calculations. Then in the leading order of the 1/N expansion
one has the following sequence of bubbles (see Fig.1)
\begin{center}
\begin{picture}(300,100)(50,-70)
\SetWidth{1.5} \Photon(0,0)(45,0){1}{6} \SetWidth{0.5}
\Text(52,0)[]{=} \Photon(60,0)(90,0){1}{6} \Text(98,0)[]{+}
\Photon(105,0)(135,0){1}{6} \ArrowArc(150,0)(15,0,180)
\ArrowArc(150,0)(15,180,360) \Photon(165,0)(195,0){1}{6}
\Text(203,0)[]{+} \Photon(210,0)(240,0){1}{6}
\ArrowArc(255,0)(15,0,180) \ArrowArc(255,0)(15,180,360)
\Photon(270,0)(300,0){1}{6} \ArrowArc(315,0)(15,0,180)
\ArrowArc(315,0)(15,180,360) \Photon(330,0)(360,0){1}{6}
\Text(373,0)[]{+...}
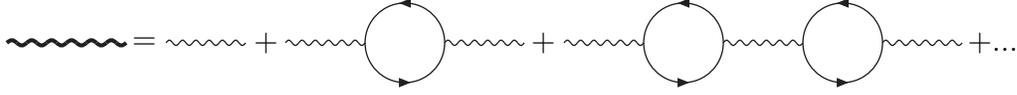
 \Text(186,-30)[]{Figure 1: The chain of diagrams giving a contribution to the
 $A$ field propagator in}
 \Text(75,-48)[]{the zeroth order of the $1/N_f$ expansion}
 \label{prop}
\end{picture}
\end{center}
summed up into a geometrical progression. The resulting photon
propagator takes the form similar to that for an auxiliary
$\sigma$ field in [I]
\begin{equation}\label{p}
  D_{\mu\nu}(p) = -\frac{i}{p^2}\left(g^{\mu\nu}-\frac{p^\mu p^\nu}{p^2}\right)
  \frac{1}{1 + e^2 f(D)(-p^2)^{D/2-2}},
\end{equation}
where
$$f(D)=\frac{\Gamma^2(D/2)\Gamma(2-D/2)}{2^{[D/2]-1}\Gamma(D)\pi^{D/2}}$$
and we put $m=0$ for simplicity.

This practically coincides with the expression obtained in scalar
theory and all the following steps just repeat those in the
latter. As was explained in [I], we change the normalization of
the gauge field $A_\mu\to A_\mu/e$ and  introduce the
dimensionless coupling $h$ associated with the triple vertex, so
the effective Lagrangian takes the form
\begin{eqnarray}\label{ef}
\nonumber  {\cal L}_{eff} &=& - \frac 14 F_{\mu\nu}\left( \frac{1}{e^2}
  + f(D)(\partial^2)^{D/2-2} (1+h)
  \right) F_{\mu\nu} - \frac {1}{2\alpha e^2} (\partial_{\mu}A_{\mu})^2 \\
&&   + i \bar{\psi_i} \hat{\partial}  \psi_i - m \bar{\psi_i} \psi_i +
\frac{\sqrt{h}}{\sqrt{N_f}} \bar{\psi_i} \hat{A} \psi_i.
\end{eqnarray}
This new dimensionless coupling $h$ enters into the gauge transformation and plays the
role of a gauge charge. The old coupling $e$, on the contrary, is dimensionful and acts
as a mass parameter in a gauge propagator. Since the coupling constant $h$ is
dimensionless the effective Lagrangian (\ref{ef}) when omitting the first term is
conformal as considered in~\cite{Anselmi} where the theory was taken in $D=3$.

Again, one has the modified Feynman rules with the photon
propagator that decreases in the Euclidean region like
$1/(p^2)^{D/2-1}$, thus improving the UV behaviour in a theory.
The only divergent graphs are those of the fermion propagator and
the triple vertex. They are both logarithmically divergent for any
odd D. The photon propagator is genuinely finite and may contain
divergencies only in subgraphs. One basically has the same graphs
as in [I] for a scalar theory but with solid lines being the
fermion ones and the dashed lines being the photon one.

The only difference (or simplification) comes from the Furry
theorem and the gauge invariance. Namely,  all triangles with
three photon external lines vanish due to the Furry theorem and
the gauge invariance which connects the fermion propagator with
the triple vertex implies that $Z_1=Z_2$. This relation holds in
the $1/N_f$ expansion like in the usual PT. Thus, using the
notation of [I], in the leading order one has
\begin{equation}
A=-\frac{\Gamma(D)(D-1)(2-D/2)}{2^{D/2} \Gamma(2-D/2) \Gamma(D/2+1) \Gamma^2(D/2)}, \
B=-A, \ C=0.
\end{equation}
The same results were obtained in~\cite{Gracey1} where the author
calculated the anomalous dimensions at the d-dimensional critical
point where the fields obey asymptotic scaling and are conformal.
This leads to the following renormalization constants in the
leading order in $1/N_f$:
\begin{eqnarray}\label{r}
Z_1=Z_2&=&1+\frac{1}{\varepsilon N_f}\frac{Ah}{(1+h)},\ \ Z_3=1
\end{eqnarray}
and, consequently, $h_B = h$. Hence, in odd-dimensional QED in the leading order of the
$1/N_f$ expansion one does not need the coupling constant renormalization; only the wave
function renormalization remains. This means that the coupling is not running.

In the second order one again has the same diagrams as in a scalar theory but with
vanishing triangles. The renormalization constant in the second order is also essentially
simplified compared to the scalar case and looks like
\begin{equation}\label{new}
Z_1=1+\frac{1}{\varepsilon N_f}\frac{A h}{1+h} +\frac{1}{\varepsilon^2 N_f^2} \frac
12\frac{A^2h^2}{(1+h)^2}+O(\frac{1}{\varepsilon N_f^2}).
\end{equation}
As explained in [I] the leading pole term $\sim 1/\varepsilon^2
N_f^2$ in (\ref{new}) can be evaluated alternatively via the pole
equations~\cite{Hooft} from the $1/\varepsilon N_f$ one with the
same result.

Like in the scalar case~[I] the original dimensionful coupling $e$
is not an expansion parameter anymore, but plays a role of a mass
and is multiplicatively logarithmically renormalized. The leading
order diagrams  are shown in Fig.2.
\begin{center}
\begin{picture}(400,120)(0,-160)

\Oval(50,-65)(15,20)(0) \ArrowArc(50,-65)(15,89,91) \ArrowArc(50,-65)(15,269,271)
\Text(50,-90)[]{a} \SetWidth{1.5} \Photon(10,-65)(30,-65){1}{3}
\Photon(70,-65)(90,-65){1}{3} \PhotonArc(50,-45)(15,-143,-37){1}{5} \SetWidth{0.5}
\Line(46,-56)(54,-64) \Line(54,-56)(46,-64)

\Oval(170,-65)(15,20)(0) \ArrowArc(170,-65)(16.5,44,46) \ArrowArc(170,-65)(16.5,134,136)
\ArrowArc(170,-65)(16.5,224,226) \ArrowArc(170,-65)(16.5,314,316) \SetWidth{1.5}
\Photon(130,-65)(150,-65){1}{3} \Photon(190,-65)(210,-65){1}{3}
\Photon(170,-50)(170,-80){1}{3} \SetWidth{0.5} \Line(166,-61)(174,-69)
\Line(174,-61)(166,-69) \Text(170,-90)[]{b}

\SetWidth{1.5} \Photon(260,-65)(280,-65){1}{3} \Photon(320,-65)(340,-65){1}{3}
\SetWidth{0.5} \ArrowLine(280,-65)(290,-55) \ArrowLine(290,-75)(280,-65)
\ArrowLine(290,-55)(290,-75) \SetWidth{1.5} \Photon(290,-55)(310,-55){1}{3}
\Photon(290,-75)(310,-75){1}{3} \SetWidth{0.5} \Line(296,-51)(304,-59)
\Line(304,-51)(296,-59) \ArrowLine(320,-65)(310,-55) \ArrowLine(310,-75)(320,-65)
\ArrowLine(310,-55)(310,-75) \Text(300,-90)[]{c}

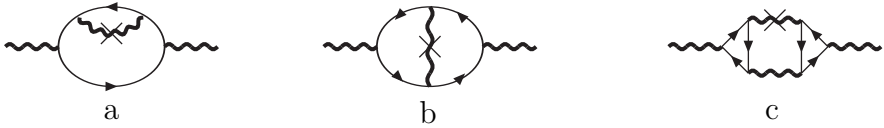
\Text(192,-110)[]{Figure 2: The first order diagrams giving a contribution
 to the $1/e^2$ renormalization}
\Text(34,-122)[]{in the $1/N_f$ expansion}
\end{picture}
\end{center}\vspace{-1cm}
They give the following contribution:
\begin{eqnarray}\label{lambda}
Diag.a &\Rightarrow& \frac{h^2}{\varepsilon N_f (1+h)^2}F,\ \
Diag.b\ \Rightarrow\ \frac{h^2}{\varepsilon N_f (1+h)^2}E,\ \
Diag.c \ \Rightarrow\ 0,  \\
&& \nonumber \\ &&\hspace*{-2.5cm}
F=\frac{\Gamma(D+1)(D/2-1)(D-1)^2(2-D/2)}{2^{D/2+1} \Gamma(2-D/2)
\Gamma(D/2+2) \Gamma^2(D/2)}, \
E=-\frac{D^2+D/2-9}{D/2(D/2-1)(D-1)}F.  \nonumber
\end{eqnarray}
So one has
\begin{equation}
Z_{1/ e^2} \ = \ 1 - \frac{1}{\varepsilon N_f} \left(
\frac{(F+E)h^2}{(1+h)^2} \right)+O(\frac{1}{N_f^2}).
\end{equation}
The solution of the RG equation for $1/e^2$ with fixed $h$ gives the  momentum dependence
of $e^2$
\begin{equation}\label{power}
\frac{1}{e^2}=\frac{1}{e^2_0}\left(\frac{p^2}{p^2_0}\right)^\gamma, \label{run}
\end{equation}
with the anomalous dimension
$$  \gamma=
\frac{\Gamma(D)(D-1)(D/2-2)(D-3)(D+2)(D-6)}{2^{D/2+1}\Gamma(D/2+2)\Gamma^2(D/2)\Gamma(2-D/2)N_f}
\frac{h^2}{(1+h)^3}. $$ The sign of $\gamma$ depends on $D$. For
$D=5,7$ $\gamma>0$, for $D=9$ $\gamma<0$ and then alternates with
every odd $D$. Notice that eq.(\ref{power}) reminds the power law
behaviour of the initial coupling in extra dimensions within the
Kaluza-Klein approach~\cite{powerbehav}.

\section{QCD}

\subsection{The effective Lagrangian}

Consider now a non-Abelian theory. Here one has some novel
features due to the presence of the triple and quartic gauge
vertices and the ghost fields. Similar to (\ref{l}) we write down
the Lagrangian for the gauge fields and $N_f$ fermions as
\begin{eqnarray}\label{ll}
  {\cal L} &=& - \frac 14 (F_{\mu \nu}^a)^2 - \frac {1}{2\alpha} (\partial_{\mu}A_{\mu}^a)^2 + i \bar{\psi_i} \hat{\partial}
  \psi_i - m \bar{\psi_i} \psi_i
  + \frac{g}{\sqrt{N_f}} \bar{\psi_i} \hat{A^a}T^a \psi_i + \partial_{\mu} \bar{c^a}
  {D}_{\mu} c_a,
  \nonumber
\end{eqnarray}
where
$$F_{\mu\nu}^a \ = \ \partial_\mu A_{\nu}^a -
\partial_\nu A_{\mu}^a + \frac{g}{\sqrt{N_f}}f^{abc}A_{\mu}^bA_{\nu}^c, \ \ \ D_\mu = \partial_\mu
+\frac{g}{\sqrt{N_f}}[A_\mu,\ ]$$
 Like in QED we choose the Landau gauge and sum up the fermion bubble diagrams into the denominator of the
gauge field propagator
\begin{equation}\label{ppp}
G^{ab}_{\mu\nu} \ = \ -\frac{i \delta^{ab}}{p^2} \frac{(g^{\mu\nu} -
\frac{p^{\mu}p^{\nu}}{p^2})}{1 + g^2 f(D)(-p^2)^{D/2-2}},
\end{equation}
where the coefficient $f(D)$ differs from the Abelian case only by
the color factor $T(R)$
$$f(D)=\frac{\Gamma^2(D/2)\Gamma(2-D/2)}{2^{[D/2]-1}\Gamma(D)\pi^{D/2}}T(R)$$
and again we put $m=0$ for simplicity.

In the non-Abelian case, contrary to the Abelian one one has the
triple and quartic self-interaction of the gauge fields. These
vertices, which are suppressed by $1/\sqrt{N_f}$ and $1/N_f$,
respectively, obtain loop corrections of the same order in
$1/N_f$. The effective vertices in the leading order are given by
the diagrams shown in Fig. 3 and 4.\vspace{-1cm}
\begin{center}
\begin{picture}(150,100)(50,-40)
\Gluon(0,0)(0,30){1}{6} \Gluon(0,0)(20,-30){1}{9} \Gluon(0,0)(-20,-30){1}{9}
\Vertex(0,0){5} \Text(25,0)[]{=} \Gluon(80,0)(80,30){1}{6} \Gluon(80,0)(50,-30){1}{9}
\Gluon(80,0)(110,-30){1}{9} \Text(115,0)[]{+} \ArrowLine(155,10)(170,-10)
\ArrowLine(170,-10)(140,-10) \ArrowLine(140,-10)(155,10) \Gluon(155,10)(155,30){1}{6}
\Gluon(170,-10)(180,-30){1}{6} \Gluon(140,-10)(130,-30){1}{6}
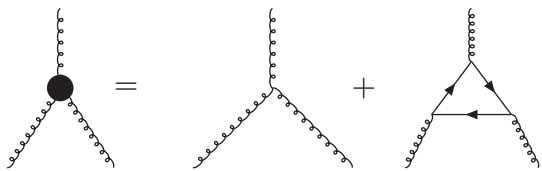
 \Text(107,-50)[]{Figure 3: The diagrams giving a contribution to the
 $A^3$ term in the zeroth order}
 \Text(-43,-68)[]{of the $1/N_f$ expansion}
 \label{prop}
\end{picture}
\end{center}

\begin{center}
\begin{picture}(150,60)(50,-70)

\Gluon(0,0)(30,-30){1}{9} \Gluon(0,0)(30,30){1}{9} \Gluon(0,0)(-30,30){1}{9}
\Gluon(0,0)(-30,-30){1}{9} \Vertex(0,0){5} \Text(35,0)[]{=} \Gluon(80,0)(50,-30){1}{9}
\Gluon(80,0)(50,30){1}{9} \Gluon(80,0)(110,30){1}{9} \Gluon(80,0)(110,-30){1}{9}
\Text(115,0)[]{+} \ArrowLine(150,15)(180,15) \ArrowLine(180,15)(180,-15)
\ArrowLine(180,-15)(150,-15) \ArrowLine(150,-15)(150,15) \Gluon(150,15)(125,30){1}{6}
\Gluon(180,15)(205,30){1}{6} \Gluon(150,-15)(125,-30){1}{6}
\Gluon(180,-15)(205,-30){1}{6}

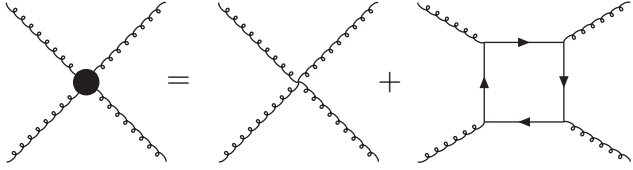
\Text(107,-50)[]{Figure 4: The diagrams giving a contribution to the $A^4$ term in the
zeroth order}
 \Text(-43,-68)[]{of the $1/N_f$ expansion}
 \label{prop}
\end{picture}
\end{center}

Thus, besides the modification of the gauge propagator one has the
modified vertices. The effective Lagrangian in the case of
vertices is not given by a simple local expression due to
complexity of the loop diagrams. So we keep it in the form of the
diagrams which have to be evaluated in integer dimension. Due to
the rules of dimensional regularization they are finite for any
odd $D$, otherwise one has to redefine them. What is crucial,
however, is that there are only three diagrams which have to be
redefined. Hence, after rescaling the gauge field $A_\mu \to
A_\mu/g$ one obtains the following effective Lagrangian:
\vspace{-1cm}
\begin{eqnarray}\label{lll}
  {\cal L}_{eff} &=& - \frac{1}{4g^2} (F_{\mu \nu}^a)^2
\begin{picture}(100,50)(-30,-5)\Text(-7,0)[]{-- (} \Gluon(0,0)(10,0){1}{3}
\Oval(20,0)(10,10)(0) \Gluon(30,0)(40,0){1}{3} \Text(55,0)[]{+} \Oval(80,0)(10,10)(0)
\Gluon(60,-10)(70,0){1}{3}\Gluon(100,-10)(90,0){1}{3}\Gluon(80,10)(80,20){1}{3}
\Text(110,0)[]{+} \Oval(140,0)(10,10)(0)
\Gluon(120,-10)(130,-5){1}{3}\Gluon(120,10)(130,5){1}{3}\Gluon(160,-10)(150,-5){1}{3}
\Gluon(160,10)(150,5){1}{3}\Text(170,0)[]{)}
\end{picture}\\
  &-&\frac {1}{2\alpha g^2} (\partial_{\mu}A_{\mu}^a)^2 + i \bar{\psi_i} \hat{\partial}
  \psi_i - m \bar{\psi_i} \psi_i
  + \frac{1}{\sqrt{N_f}} \bar{\psi_i} \hat{A^a}T^a \psi_i + \partial_{\mu} \bar{c^a}
  {D}_{\mu} c_a,
  \nonumber
\end{eqnarray}
Notice that dimensionful coupling $g$ drops from all terms except for the first one and
is not an expansion parameter anymore.

Calculating the degree of divergence after summing up the diagrams of the zeroth order,
similar to the scalar case and QED, one has only four types of logarithmically divergent
diagrams: the fermion and the ghost propagators, the fermion-gauge-vertex and
ghost-gauge-ghost vertex. The gauge propagator as well as pure gauge vertices are finite
and may contain only divergent subgraphs.

The next step is the introduction of a dimensionless coupling $h$.
Here one should be accurate since this coupling enters not only
into the triple gauge-fermion vertex, but due to the gauge
invariance should be present in gauge and gauge-ghost vertices. It
should be the same in all three of them. In the case of a gauge
theory, the coupling $h$ enters the gauge transformation and acts
as a gauge charge of the fermion and gauge fields.

When constructing the Feynman diagrams, one reproduces the
one-loop cycles that are already present in the effective
Lagrangian (\ref{lll}) but with additional factors h. In the
scalar or QED case, this happened only  for the propagator, but
here it is also true for the vertices. As a result, the final
expression for the effective Lagrangian takes the form
\vspace{-1cm}
\begin{eqnarray}\label{lll}
  {\cal L}_{eff} &=& - \frac{1}{4g^2} (F_{\mu \nu}^a)^2
\begin{picture}(100,50)(-30,-5)\Text(-7,0)[]{-- (} \Gluon(0,0)(10,0){1}{3}
\Oval(20,0)(10,10)(0) \Gluon(30,0)(40,0){1}{3} \Text(55,0)[]{+} \Oval(80,0)(10,10)(0)
\Gluon(60,-10)(70,0){1}{3}\Gluon(100,-10)(90,0){1}{3}\Gluon(80,10)(80,20){1}{3}
\Text(110,0)[]{+} \Oval(140,0)(10,10)(0)
\Gluon(120,-10)(130,-5){1}{3}\Gluon(120,10)(130,5){1}{3}\Gluon(160,-10)(150,-5){1}{3}
\Gluon(160,10)(150,5){1}{3}\Text(185,0)[]{)\ (1+h)}
\end{picture}\\
  &-&\frac {1}{2\alpha g^2} (\partial_{\mu}A_{\mu}^a)^2 + i \bar{\psi_i} \hat{\partial}
  \psi_i - m \bar{\psi_i} \psi_i
  + \frac{h}{\sqrt{N_f}} \bar{\psi_i} \hat{A^a}T^a \psi_i + \partial_{\mu} \bar{c^a}
  {D}_{\mu} c_a,
  \nonumber
\end{eqnarray}
where
$$F_{\mu\nu}^a \ = \ \partial_\mu A_{\nu}^a - \partial_\nu A_{\mu}^a + \frac{\sqrt{h}}{\sqrt{N_f}}f^{abc}A_{\mu}^bA_{\nu}^c \
\ ,\ \ D_{\mu}c_a \ = \
\partial_{\mu}c_a+\frac{\sqrt{h}}{\sqrt{N_f}}f^{abc}A^b_{\mu}c^c.$$

\subsection{Properties of the $1/N_f$ expansion}

Consider now the leading order calculations. We start with the
$1/N_f$ terms for the fermion and the triple fermion-gauge-fermion
vertex. The diagrams are shown in Fig.5. The first two are the
same as in QED. The third diagram contains new effective vertex
which includes the usual triple vertex and the fermion triangle.
The usual vertex does not give a contribution since it is finite
by a simple power counting. At the same time, the fermion triangle
is momentum dependent and the resulting diagram is logarithmically
divergent.
\begin{center}
\begin{picture}(250,130)(0,-55)
\ArrowLine(-10,0)(10,0) \ArrowLine(10,0)(40,0) \ArrowLine(40,0)(60,0) \SetWidth{1.5}
\GlueArc(25,0)(15,0,180){1}{6} \SetWidth{0.5} \Text(25,-10)[]{a}

\SetWidth{1.5} \Gluon(110,55)(110,35){1}{3} \SetWidth{0.5} \ArrowLine(80,-0)(110,35)
\ArrowLine(110,35)(140,-0) \SetWidth{1.5} \Gluon(89,10)(131,10){1}{6} \SetWidth{0.5}
\Text(110,-10)[]{b}

\ArrowLine(170,0)(190,0) \ArrowLine(190,0)(230,0) \ArrowLine(230,0)(250,0) \SetWidth{1.5}
\Gluon(190,0)(210,35){1}{6} \Gluon(230,0)(210,35){1}{6}
\Vertex(210,35){3} \Gluon(210,35)(210,60){1}{3} \SetWidth{0.5} \Text(210,-10)[]{c}
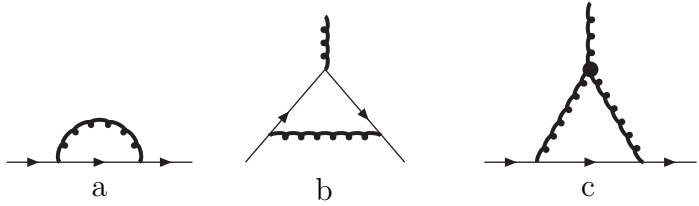
\Text(119,-30)[]{Figure 5: The leading order diagrams giving a contribution to the $\psi$
field propagator} \Text(3,-43)[]{and the triple vertex  in $1/N_f$ expansion}
\end{picture}
\end{center}

Calculating the singular parts of the diagrams of Fig.5 in
dimensional regularization with $D'=D-2\varepsilon$ one finds
\begin{eqnarray}\label{sing_ferm}
Diag.a &\Rightarrow& \frac{1}{\varepsilon N_f} \frac{h}{1+h} A,\ \
\ Diag.b\ \Rightarrow\ \frac{1}{\varepsilon N_f} \frac{h}{1+h} B,\
\ \
Diag.c \ \Rightarrow\ \frac{1}{\varepsilon N_f} \frac{h}{1+h} C ,\\
&& \nonumber \\ &&\hspace*{-2.5cm} A=-\frac{\Gamma(D)(D-1)(2-D/2)C_F}{2^{D/2}
\Gamma(2-D/2) \Gamma(D/2+1) \Gamma^2(D/2)T} , \ B=-\frac{C_F-C_A/2}{C_F}A, \
C=-\frac{(1-D/2)C_A}{2(2-D/2)C_F}A \nonumber,
\end{eqnarray}
which is again in agreement with~\cite{Gracey2}.
Notice that the third diagram is  proportional to  $h/(1+h)$ instead of $h^2/(1+h)^2$ as
in the scalar case. The reason is that now we have an effective triple gauge vertex
proportional to $\sqrt{h}(1+h)$ instead of $\sqrt{h}$ that cancels one factor of
$h/(1+h)$.

Therefore, in the leading order in the $1/N_f$ expansion the renormalization constants
take the form
\begin{eqnarray}\label{r}
Z_2^{-1}&=&1-\frac{1}{\varepsilon N_f}\frac{Ah}{(1+h)},\\
Z_1&=&1-\frac{1}{\varepsilon N_f}\frac{(B+C)h}{(1+h)}, \\
Z_h&=& Z_1^2Z_2^{-2}=1-\frac{1}{\varepsilon N_f
}\frac{2(A+B+C)h}{(1+h)}.
\end{eqnarray}

To check the gauge invariance, we calculated the renormalization
of the coupling through the gauge-ghost interaction.  The leading
diagrams are shown in Fig.6.

\begin{center}
\begin{picture}(250,130)(0,-55)
\DashArrowLine(-10,0)(10,0){2} \DashArrowLine(10,0)(40,0){2}
\DashArrowLine(40,0)(60,0){2} \SetWidth{1.5} \GlueArc(25,0)(15,0,180){1}{6}
\SetWidth{0.5} \Text(25,-10)[]{a}

\SetWidth{1.5} \Gluon(110,55)(110,35){1}{3} \SetWidth{0.5}
\DashArrowLine(80,-0)(110,35){2} \DashArrowLine(110,35)(140,-0){2} \SetWidth{1.5}
\Gluon(89,10)(131,10){1}{6} \SetWidth{0.5} \Text(110,-10)[]{b}

\DashArrowLine(170,0)(190,0){2} \DashArrowLine(190,0)(230,0){2}
\DashArrowLine(230,0)(250,0){2} \SetWidth{1.5} \Gluon(190,0)(210,35){1}{6}
\Gluon(230,0)(210,35){1}{6}
\Vertex(210,35){3} \Gluon(210,35)(210,60){1}{3} \SetWidth{0.5} \Text(210,-10)[]{c}
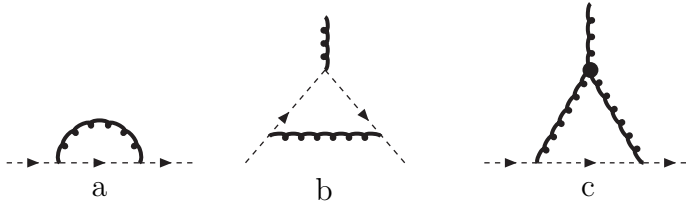
\Text(123,-30)[]{Figure 6: The leading order diagrams giving a contribution to the ghost
field propagator} \Text(3,-43)[]{and the triple vertex  in $1/N_f$ expansion}
\end{picture}
\end{center}
Calculating the singular parts of the diagrams  in dimensional regularization one finds
\begin{eqnarray}\label{sing_ghost}
Diag.a &\Rightarrow& \frac{1}{\varepsilon N_f} \frac{h}{1+h}A',\ \
Diag.b\ \Rightarrow\ \frac{1}{\varepsilon N_f} \frac{h}{1+h} B',\
\ Diag.c \ \Rightarrow\ \frac{1}{\varepsilon N_f} \frac{h}{1+h} C' ,  \nonumber\\
&& \nonumber \\ &&\hspace*{-2.5cm} A'=-\frac{\Gamma(D)(D-1)C_A}{2^{D/2+1} \Gamma(2-D/2)
\Gamma(D/2+1) \Gamma^2(D/2)T} , \ \ \ B'=0, \ \ C'=0,
\end{eqnarray}
which gives the following renormalization constants in the ghost sector
\begin{eqnarray}\label{ghost}
\widetilde{Z}_1&=&1,  \\
\widetilde{Z}_2^{-1}&=&1-\frac{1}{\varepsilon N_f}\frac{A'h}{1+h},\\
Z_h&=&\widetilde{Z}_1^2\widetilde{Z}_2^{-2}=1-\frac{2}{\varepsilon
N_f}\frac{A'h}{1+h}.
\end{eqnarray}
One can see that the following relation holds:
\begin{equation}\label{unit}
 A+B+C \ = \ A'+B'+C',
\end{equation}
which follows from the gauge invariance.

We look now at the next-to-leading order to compare it with the
scalar case. The corresponding diagrams for the fermion propagator
are shown in Fig.7. They require some explanation. The first line
of diagrams in Fig.7 is obtained from the one-loop diagrams of
Fig.5 by inserting into the vertex or the fermion line of the
one-loop divergent subgraphs from Fig.5. For example, the diagram
$d$ in Fig.7 is the diagram $a$ from Fig.5 with divergent one-loop
subgraph $c$ from Fig.5 substituted instead of the initial vertex.
The second line of the diagrams in Fig.7 is obtained from the
"forbidden" diagram of Fig.8 by inserting the same one-loop
divergent subgraphs from Fig.5. The diagram $e$ is the diagram of
Fig.8 with insertion of the subgraph $a$ from Fig.5 into the
fermion line (see Fig.9) and the diagram $g$  comes from the
insertion of the subgraph $c$ from Fig.5 instead of one of the
vertices in the fermion loop (see Fig.10).


\begin{center}
\begin{picture}(400,235)(0,-170)

\ArrowLine(0,0)(10,0) \ArrowLine(10,0)(20,0)\ArrowLine(20,0)(70,0)
\ArrowLine(60,0)(70,0)\ArrowLine(70,0)(80,0) \SetWidth{1.5}
\GlueArc(40,0)(20,0,180){1}{9} \GlueArc(40,0)(30,0,180){1}{11} \SetWidth{0.5}
\Text(40,-30)[]{a}

\ArrowLine(100,0)(110,0) \ArrowLine(110,0)(130,0) \ArrowLine(130,0)(150,0)
\ArrowLine(150,0)(170,0) \ArrowLine(170,0)(180,0) \SetWidth{1.5}
\GlueArc(130,0)(20,0,180){1}{7} \GlueArc(150,0)(20,180,360){1}{7} \SetWidth{0.5}
\Text(140,-30)[]{b}

\ArrowLine(200,0)(210,0) \ArrowLine(210,0)(230,0) \ArrowLine(230,0)(250,0)
\ArrowLine(250,0)(270,0) \ArrowLine(270,0)(280,0) \SetWidth{1.5}
\GlueArc(220,0)(10,0,180){1}{6} \GlueArc(260,0)(10,0,180){1}{6} \SetWidth{0.5}
\Text(240,-30)[]{c}

\ArrowLine(300,0)(320,0) \ArrowLine(320,0)(340,0) \ArrowLine(340,0)(360,0)
\ArrowLine(360,0)(380,0) \SetWidth{1.5} \Gluon(320,0)(340,40){1}{6}
\Gluon(340,0)(340,40){1}{5} \Gluon(360,0)(340,40){1}{6} \Vertex(340,40){3} \SetWidth{0.5}
\Text(340,-30)[]{d}

\ArrowLine(0,-100)(10,-100) \ArrowLine(10,-100)(90,-100) \ArrowLine(90,-100)(100,-100)
\Oval(50,-65)(15,20)(0) \ArrowArc(50,-65)(15,269,271) \ArrowArc(50,-65)(15,89,91)
\SetWidth{1.5} \GlueArc(50,-100)(40,120,180){1}{5} \GlueArc(50,-100)(40,0,60){1}{5}
\GlueArc(50,-45)(15,-143,-37){1}{4} \SetWidth{0.5} \Text(50,-120)[]{e}

\ArrowLine(120,-100)(130,-100) \ArrowLine(130,-100)(210,-100)
\ArrowLine(210,-100)(220,-100) \Oval(170,-65)(15,20)(0) \ArrowArc(170,-65)(16.5,134,136)
\ArrowArc(170,-65)(16.5,44,46) \ArrowArc(170,-65)(16.5,224,226)
\ArrowArc(170,-65)(16.5,314,316) \SetWidth{1.5} \GlueArc(170,-100)(40,120,180){1}{5}
\GlueArc(170,-100)(40,0,60){1}{5} \Gluon(170,-50)(170,-80){1}{5} \SetWidth{0.5}
\Text(170,-120)[]{f}

\ArrowLine(250,-100)(260,-100) \ArrowLine(260,-100)(340,-100)
\ArrowLine(340,-100)(350,-100) \SetWidth{1.5} \GlueArc(280,-100)(20,90,180){1}{5}
\GlueArc(320,-100)(20,0,90){1}{5}
\Gluon(280,-80)(310,-70){1}{5} \Gluon(280,-80)(310,-90){1}{5} \Vertex(280,-80){3}
\SetWidth{0.5} \ArrowLine(320,-80)(310,-70) \ArrowLine(310,-90)(320,-80)
\ArrowLine(310,-70)(310,-90) \Text(300,-120)[]{g}

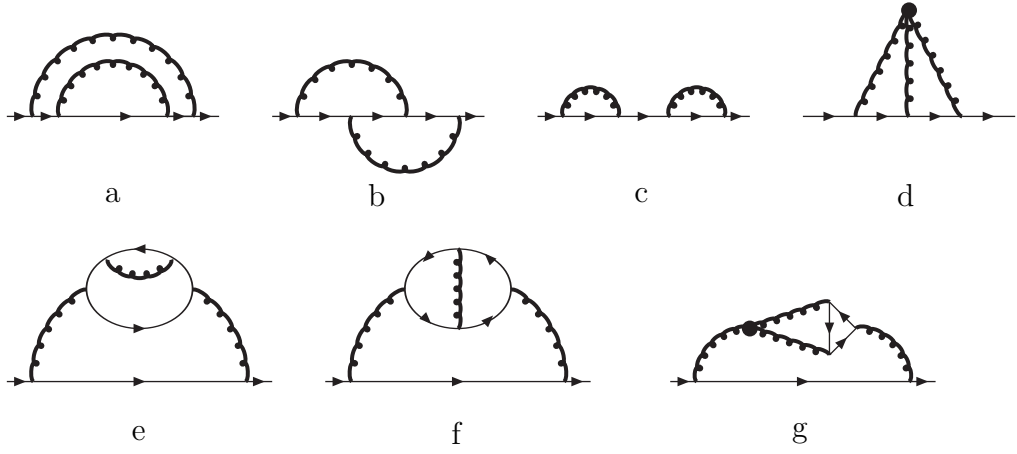
\Text(194,-140)[]{Figure 7: The second order diagrams giving a
contribution to the fermion propagator} \Text(31,-152)[]{in the
$1/N_f$ expansion}
\end{picture}
\end{center}

\begin{center}
\begin{picture}(110,95)(0,-35)
\ArrowLine(-10,0)(10,0) \ArrowLine(10,0)(90,0) \ArrowLine(90,0)(110,0)
\Oval(50,35)(15,20)(0) \ArrowArc(50,35)(15,89,91) \ArrowArc(50,35)(15,269,271)
\SetWidth{1.5} \GlueArc(50,0)(40,120,180){1}{7} \GlueArc(50,0)(40,0,60){1}{7}
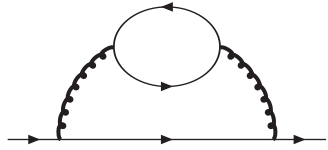
\SetWidth{0.5} \Text(50,-20)[]{Figure 8: The "forbidden" loop diagram}
\end{picture}
\end{center}

\begin{center}
\begin{picture}(270,95)(0,-35)
\ArrowLine(-10,0)(10,0) \ArrowLine(10,0)(90,0) \ArrowLine(90,0)(110,0)
\Oval(50,35)(15,20)(0) \ArrowArc(50,35)(15,89,91) \ArrowArc(50,35)(15,269,271)
\SetWidth{1.5} \GlueArc(50,0)(40,120,180){1}{7} \GlueArc(50,0)(40,0,60){1}{7}
\SetWidth{0.5}

\Text(112,20)[]{+}

\ArrowLine(120,10)(160,10)\SetWidth{1.5} \GlueArc(140,10)(15,0,180){1}{7} \SetWidth{0.5}

\Text(165,20)[]{$\rightarrow$}

\ArrowLine(170,0)(180,0) \ArrowLine(180,0)(260,0) \ArrowLine(260,0)(270,0)
\Oval(220,35)(15,20)(0) \ArrowArc(220,35)(15,269,271) \ArrowArc(220,35)(15,89,91)
\SetWidth{1.5} \GlueArc(220,0)(40,120,180){1}{5} \GlueArc(220,0)(40,0,60){1}{5}
\GlueArc(220,55)(15,-143,-37){1}{4} \SetWidth{0.5}

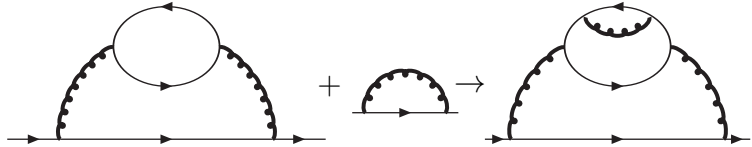
\Text(133,-20)[]{Figure 9: The diagram $e$ from Fig.7 as a result
of insertion of the diagram $a$ from Fig.5} \Text(-37,-35)[]{into
the fermion line.}
\end{picture}
\end{center}

\begin{center}
\begin{picture}(270,95)(0,-35)
\ArrowLine(-10,0)(10,0) \ArrowLine(10,0)(90,0) \ArrowLine(90,0)(110,0)
\Oval(50,35)(15,20)(0) \ArrowArc(50,35)(15,89,91) \ArrowArc(50,35)(15,269,271)
\SetWidth{1.5} \GlueArc(50,0)(40,120,180){1}{7} \GlueArc(50,0)(40,0,60){1}{7}
\SetWidth{0.5}

\Text(112,20)[]{+}

\ArrowLine(120,0)(160,0) \ArrowLine(130,20)(150,20) \ArrowLine(150,20)(140,30)
\ArrowLine(140,30)(130,20) \SetWidth{1.5} \Gluon(130,0)(130,20){1}{5}
\Gluon(150,0)(150,20){1}{5} \Gluon(140,30)(140,50){1}{5} \SetWidth{0.5}

\Text(165,20)[]{$\rightarrow$}

\ArrowLine(170,0)(180,0) \ArrowLine(180,0)(260,0) \ArrowLine(260,0)(270,0) \SetWidth{1.5}
\GlueArc(200,0)(20,90,180){1}{5} \GlueArc(240,0)(20,0,90){1}{5}
\Gluon(200,20)(230,30){1}{5} \Gluon(200,20)(230,10){1}{5} \Vertex(200,20){3}
\SetWidth{0.5} \ArrowLine(240,20)(230,30) \ArrowLine(230,10)(240,20)
\ArrowLine(230,30)(230,10)

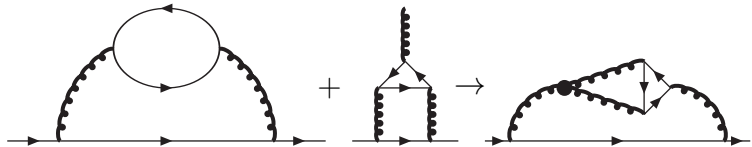
\Text(133,-20)[]{Figure 10: The diagram $g$ from Fig.7 as a result of insertion of the
diagram $c$ from Fig.5} \Text(50,-35)[]{instead of one of the vertices in the diagram
from Fig.8.}
\end{picture}
\end{center}
%

All these diagrams are double logarithmically divergent, i.e.,
contain both single and double poles in dimensional
regularization. We calculated the leading double poles after
subtraction of the divergent subgraphs, i.e., performed  the
$R'$-operation. The answer is:
\begin{eqnarray}\label{sing2}
Diag.a &\Rightarrow& -\frac{1}{\varepsilon^2
N_f^2}\frac{A^2h^2}{2(1+h)^2},\  Diag.b\ \Rightarrow\
-\frac{1}{\varepsilon^2 N_f^2}\frac{ABh^2}{(1+h)^2},\
Diag.c \ \Rightarrow\ -\frac{1}{\varepsilon^2 N_f^2}\frac{A^2h^2}{(1+h)^2}, \nonumber \\
Diag.d &\Rightarrow& -\frac{1}{\varepsilon^2
N_f^2}\frac{ACh^2}{(1+h)^2},\ \ \ Diag.e\ \Rightarrow
\frac{1}{\varepsilon^2 N_f^2}\frac{2A^2h^3}{3(1+h)^3},\
\\ Diag.f &\Rightarrow& \frac{1}{\varepsilon^2 N_f^2}\frac{2ABh^3}{3(1+h)^3},
\ \ \ Diag.g  \Rightarrow \frac{1}{\varepsilon^2
N_f^2}\frac{2ACh^3}{3(1+h)^3}.\nonumber
\end{eqnarray}
We performed also the calculation for the fermion-gauge-fermion
vertex but do not present the diagram-by-diagram result because of
the lack of space and give only the final answer.

Having all this in mind we come to the final expressions for the Z factors in the second
order of the $1/N_f$ expansion in the fermion sector:
\begin{eqnarray}\label{newg}
Z_1&=&1-\frac{1}{\varepsilon N_f}\frac{(B+C)h}{1+h} +
\frac{1}{\varepsilon^2 N_f^2}\left( \frac
32\frac{(B+C)^2h^2}{(1+h)^2} +
\frac{A(B+C)h^2}{(1+h)^2}\right.\nonumber
\\ &&\left. -\frac {2}{3}\frac{(B+C)^2h^3}{(1+h)^3} - \frac
23\frac{A(B+C)h^3}{(1+h)^3} \right)
+O(\frac{1}{\varepsilon N_f^2}), \\
Z_2^{-1}&=&1-\frac{1}{\varepsilon
N_f}\frac{Ah}{1+h}+\frac{1}{\varepsilon^2 N_f^2}\left( \frac
32\frac{A^2h^2}{(1+h)^2} + \frac{A(B+C)h^2}{(1+h)^2} \right.\nonumber \\
&&\left. - \frac 23\frac{A(A+B+C)h^3}{(1+h)^3}
\right)+O(\frac{1}{\varepsilon N_f^2}). \label{new2g}
\end{eqnarray}
The same calculation in  the ghost sector gives
\begin{eqnarray}\label{newghost}
\widetilde{Z}_1&=&1 ,  \\
\widetilde{Z}_2^{-1}&=&1-\frac{1}{\varepsilon
N_f}\frac{A'h}{1+h}+\frac{1}{\varepsilon^2 N_f^2}\left( \frac
32\frac{A'^2h^2}{(1+h)^2}
- \frac 23\frac{A'(A+B+C)h^3}{(1+h)^3}
\right)+O(\frac{1}{\varepsilon N_f^2}). \label{new2ghost}
\end{eqnarray}
Notice the absence of the ghost-gauge-ghost vertex renormalization.

The final second order expression for the coupling renormalization calculated in both
ways having in mind relation (\ref{unit}) is
\begin{equation}\label{Renorm}
Z_h=1 - \frac {1}{\varepsilon N_f} \frac{2(A+B+C)h}{1+h} +
\frac{1}{\varepsilon^2 N_f^2}
  \left(4\frac{(A+B+C)^2h^2}{(1+h)^2} -
\frac43\frac{(A+B+C)^2h^3}{(1+h)^3}\right)+O(\frac{1}{\varepsilon
N_f^2}).
\end{equation}

Like in the scalar and QED case, one can also calculate the
renormalization of the original coupling $g^2$. The leading order
diagrams are shown in Fig. 11 which give the following singular
parts like in~\cite{Gracey2}
\begin{center}
\begin{picture}(400,120)(0,-160)

\Oval(50,-65)(15,20)(0) \ArrowArc(50,-65)(15,89,91) \ArrowArc(50,-65)(15,269,271)
\Text(50,-95)[]{a} \SetWidth{1.5} \Gluon(10,-65)(30,-65){1}{3}
\Gluon(70,-65)(90,-65){1}{3} \GlueArc(50,-45)(15,-143,-37){1}{5} \SetWidth{0.5}
\Line(46,-56)(54,-64) \Line(54,-56)(46,-64)

\Oval(160,-65)(15,20)(0) \ArrowArc(160,-65)(16.5,44,46) \ArrowArc(160,-65)(16.5,134,136)
\ArrowArc(160,-65)(16.5,224,226) \ArrowArc(160,-65)(16.5,314,316) \SetWidth{1.5}
\Gluon(120,-65)(140,-65){1}{3} \Gluon(180,-65)(200,-65){1}{3}
\Gluon(160,-50)(160,-80){1}{3} \SetWidth{0.5} \Line(156,-61)(164,-69)
\Line(164,-61)(156,-69) \Text(160,-95)[]{b}

\SetWidth{1.5} \Gluon(220,-65)(240,-65){1}{3} \Gluon(280,-65)(300,-65){1}{3}
\GlueArc(260,-65)(20,0,180){1}{7} \GlueArc(260,-65)(20,180,360){1}{7} \Vertex(240,-65){3}
\Vertex(280,-65){3} \SetWidth{0.5} \Line(256,-39)(264,-49) \Line(264,-39)(256,-49)
\Text(260,-95)[]{c}

\SetWidth{1.5} \Gluon(320,-65)(340,-65){1}{3} \Gluon(380,-65)(400,-65){1}{3}
\GlueArc(360,-65)(20,0,180){1}{7} \GlueArc(360,-65)(20,180,360){1}{7} \Vertex(380,-65){3}
\SetWidth{0.5} \Line(346,-60)(334,-70) \Line(334,-60)(346,-70) \Text(360,-95)[]{d}

\Text(192,-115)[]{Figure 11: The first order diagrams giving a contribution to the
$1/g^2$ renormalization} \Text(22,-130)[]{in $1/N_f$ expansion}
\end{picture}
\end{center}\vspace{-1cm}

\begin{eqnarray}\label{lambda}
\nonumber Diag.a &\Rightarrow& \frac{1}{\varepsilon N_f}
\frac{h^2}{(1+h)^2}F,\ \ Diag.b\ \Rightarrow\ \frac{1}{\varepsilon
N_f} \frac{h^2}{(1+h)^2}E,  \\ && \hspace*{-2.5cm} Diag.c\
\Rightarrow\
\frac{1}{\varepsilon N_f} \frac{h^2}{(1+h)^2}G, \ \ \ Diag.d\ \Rightarrow\ \frac{1}{\varepsilon N_f} \frac{h^2}{(1+h)^2}H,  \\
&& \nonumber \hspace*{-2.5cm}
F=\frac{\Gamma(D+1)(D/2-1)(D-1)^2(2-D/2)C_F}{2^{D/2+1}
\Gamma(2-D/2) \Gamma(D/2+2) \Gamma^2(D/2)T(R)}, \
E=-\frac{D^2+D/2-9}{D/2(D/2-1)(D-1)}\frac{C_F-C_A/2}{C_F}F, \\
&& \nonumber \hspace*{-2.5cm}
G=\frac{4(D/2)^6-6(D/2)^5+18(D/2)^4-67(D/2)^3+85(D/2)^2-19D+6}{2(D-1)^2(1-D/2)^2(2-D/2)D}
\frac{C_A}{C_F} F, \\
&& \nonumber \hspace*{-2.5cm}
H=\frac{D^3-D^2/2-2D+1}{D(1-D/2)(2-D/2)(D-1)^2}\frac{C_A}{C_F }F.
\nonumber
\end{eqnarray}
The corresponding renormalization constant looks like
\begin{equation}
Z_{1/ g^2} \ = \ 1 - \frac{1}{\varepsilon N_f}
\frac{(F+E+G+H)h^2}{(1+h)^2}.
\end{equation}

\subsection{Renormalization group in $1/N_f$ expansion}

Having these expressions for the Z factors one can construct the coupling constant
renormalization and the  corresponding RG functions. One has as usual in dimensional
regularization
\begin{eqnarray}\label{rg}
  h_B&=&(\mu^2)^\varepsilon hZ_1^2Z_2^{-2}=(\mu^2)^\varepsilon \left(
  h+\sum_{n=1}^{\infty}\frac{a_n(h,N_f)}{\varepsilon^n}\right),\\
  Z_i&=&1+\sum_{n=1}^{\infty}\frac{c^i_n(h,N_f)}{\varepsilon^n},\label{zz}
\end{eqnarray}
where the first coefficients $a_n$ and $c^i_n$ can be deduced from
eqs.(\ref{newg},\ref{new2g},\ref{newghost},\ref{new2ghost}).

This allows one to get the anomalous dimensions and the beta function defined as
\begin{eqnarray}\label{anom}
  \gamma(h,N_f)&=&-\mu^2\frac{d}{d\mu^2}\log Z = h\frac{d}{dh}c_1, \\
 \beta(h,N_f)&=&2h(\gamma_1+\gamma_2)=(h\frac{d}{dh}-1)a_1.
\end{eqnarray}
With the help of eqs.(\ref{newg},\ref{new2g}) one gets in the leading order of the
$1/N_f$ expansion
\begin{eqnarray}\label{dim}
  \gamma_2(h,N_f)&=&-\frac {1}{N_f} \frac{Ah}{(1+h)^2}, \ \ \
  \gamma_1(h,N_f)=-\frac {1}{N_f} \frac{(B+C)h}{(1+h)^2},\\
  \widetilde{\gamma}_2(h,N_f)&=&-\frac {1}{N_f} \frac{A'h}{(1+h)^2}, \ \ \
  \widetilde{\gamma}_1(h,N_f)=O(\frac{1}{N_f^2}),\\
  \beta(h,N_f)&=&-\frac {1}{N_f} \frac{2(A+B+C)h^2}{(1+h)^2}, \label{betaferm}
\end{eqnarray}

It is instructive to check the pole equations~\cite{Hooft} that express the coefficients
of the higher order poles in $\varepsilon$ of the Z factors via the coefficients of a
simple pole. For $Z_2^{-1}$ one has, according to (\ref{new2g}),
\begin{eqnarray}\label{polesgauge}
  c_1(h,N_f)&=&-\frac {1}{N_f} \frac{Ah}{1+h}, \\
  c_2(h,N_f)&=& \frac{1}{N_f^2}
  \left(
\frac 32\frac{A^2h^2}{(1+h)^2}+\frac{A(B+C)h^2}{(1+h)^2}-\frac
23\frac{A(A+B+C)h^3}{(1+h)^3}\right).\label{polesgauge2}
\end{eqnarray}
At the same time, the coefficient $c_2$ can be expressed through
$c_1$ via the pole equations as
\begin{equation}\label{p2g}
  h\frac{dc_2}{dh}=\gamma_2c_1+\beta\frac{dc_1}{dh},
\end{equation}
Integrating this equation one gets for $c_2$ the expression coinciding with
(\ref{polesgauge2}) which was obtained by direct diagram evaluation.

We have also checked  the pole equations for the renormalized coupling. From
eq.(\ref{Renorm}) one gets the coefficients of the coupling  constant renormalization
factor $Z_h$
\begin{eqnarray}\label{poles}
  a_1(h,N_f)&=&-\frac {1}{N_f} \frac{2(A+B+C)h^2}{1+h} , \\
  a_2(h,N_f)&=& \frac{1}{N_f^2}
  \left(
4\frac{(A+B+C)^2h^3}{(1+h)^2} - \frac43\frac{(A+B+C)^2h^4}{(1+h)^3}\right).\label{poles3}
\end{eqnarray}
At the same time, from the pole equations one has
\begin{equation}\label{rg}
(h\frac{d}{dh}-1)a_n=\beta\frac{da_{n-1}}{dh}.
\end{equation}
The coefficient $a_2$ evaluated in this way coincides with
(\ref{poles3}).

Equation (\ref{Renorm}) gives us the sign of the beta function. In the leading order one
has
\begin{equation}\label{beta}
\frac{dh}{dt}=\beta(h)= \frac{\Gamma(D)(D-1)C_A}{2^{D/2} \Gamma(2-D/2) \Gamma(D/2+1)
\Gamma^2(D/2)N_fT}\frac{h^2}{(1+h)^2},
\end{equation}
which means that $\beta(h)<0$ for $D=5$,  $\beta(h)>0$ for $D=7$ and then alternates with
$D$.

Notice that in QCD, contrary to QED, all Feynman diagrams contain group factors so that
the actual expansion parameter becomes $N_c/N_f$, thus requiring that this ratio is
small. Of course, in non-Abelian theories the $1/N_c$ expansion would be preferable,
since it accumulates the interactions of the gauge fields, however, in this case already
the lowest approximation consists of all planar diagrams and is not known~\cite{Hooft1}.

\section{Conclusion}

We conclude that in higher dimensional gauge theories like in the
scalar case despite formal non-renormalizability it is possible to
construct renormalizable $1/N_f$ expansion which obeys all the
rules of a usual perturbation theory. The expansion parameter is
dimensionless, the coupling is running logarithmically, all
divergencies are absorbed into the renormalization of the wave
function and the coupling.  The original dimensionful coupling
plays a role of mass and is renormalized multiplicatively.
Expansion over this coupling is singular and creates the usual
nonrenormalizable terms.

Properties of the $1/N_f$ expansion do not depend on the
space-time dimension if it is odd. In even dimension our formulas
after subtraction contain a logarithm which creates some technical
problems in calculations.  We plan to consider these theories
later.

There is one essential point that we omitted in our discussion,
namely, the analytical properties of the gauge field propagator
and the unitarity of a resulting theory. As one can see from
eqs.(\ref{p}) and (\ref{ppp}), the propagator of the gauge field
contains a cut starting from $p^2=0$ in the massless case or from
$4m^2$ in the massive one. It may also contain poles in complex
momentum plane. This imposes the question about unitarity of a
resulting theory.

The unitarity of such an expansion was shown in~\cite{Arefeva} at
the tree level.  In the case of loops, one can show that  all the
cuts imposed on diagrams when applying Cutkosky rules~\cite{Cut}
in any order of perturbation theory lead to the usual asymptotic
states on the mass shell and no new states appear. From this point
of view the theory remains unitary in physical space. The
scattering amplitudes also behave safely decreasing with momenta,
thus do not violating the unitarity requirement.

The only source of conceptual problem is the presence of poles in
the propagator in the complex momentum plane. They may appear in
the Euclidean region for negative $k^2$ (as in D=5) or for complex
$k^2$ (as in D=7). In the first case, they create a problem when
integrating over Euclidean momenta and one might take the
integrals in a sense of a principle value, while in the second
case this does not happen, but in both the cases the presence of
this type of poles might be interpreted as appearance of new
(ghost) states. Note, however, that the form of a dressed
propagator is not specific for the $1/N_f$ expansion, but is
typical of any perturbation theory. For instance, in the usual QED
or any other renormalizable theory the dressed propagator has the
same form and may obtain poles in the momentum plane (the Landau
pole is an example). This is incompatible with the
K\"allen-Lehmann representation leading to a non-positive spectral
function~\cite{Kr} or to a negative contribution from pole terms,
which signals negative norm states exist.

A usual way to avoid this problem is to consider the effective
theory at low momenta below the Landau pole assuming that the
situation is improved in a proper higher dimensional theory.
Indeed, in the four dimensional QED the Landau pole exists at the
energies above the Planck scale and one can safely use QED at
lower momenta. In our case, however, due to power like behaviour
it happens much earlier and severely constrains the region of
validity of this theory.

There are several attempts to build renormalizable effective
quantum gravity using a kind of $1/N$ expansion~\cite{gravity},
where the role of an expansion parameter $1/N$ is played by the
number space-time dimensions. The large $D$ limit in this case is
very similar to the large $N_c$  planar diagram limit in the
Yang-Mills theory considered by 't Hooft~\cite{Hooft1}. The
technique similar to the $1/N_f$ expansion is used also in
\cite{Ward}, though no large parameter appears. The author, as
in~\cite{YFS}, sums up the soft graviton corrections to the
propagators of the scalar field  to get an improved propagator
which decreases faster than any power of $|k^2|$. Though this
partial resummation is similar to our $1/N$ expansion, the absence
of an expansion parameter does not justify, to our mind,  the
selected set of diagrams. From this point of view the $1/N$
expansion is more consistent and contains the guiding line for
such a selection.

\section*{Acknowledgements}
Financial support from RFBR grant \# 05-02-17603 and grant of the
Ministry of Education and Science of the Russian Federation \#
5362.2006.2 is kindly acknowledged. We are grateful to I.Aref'eva,
G.Efimov, A.Kotikov, N.Krasnikov, S.Mikhailov and A.Sheplyakov for
valuable discussions.


\begin{thebibliography}{99}
\bibitem{KVI} D.I.Kazakov and G.S.Vartanov, arXiv:hep-th/0607177.
\bibitem{Justin} M.Moshe, J.Zinn-Justin, Phys.Rept., {\bf 385} (2003)
69-228 [arXiv:hep-th/0306133]; \\
J.Zinn-Justin, Phys.Rept., SACLAY-SPH-T-97-018
[arXiv:hep-th/9810198].
\bibitem{Hooft0} G.t'Hooft, SPIN-2002-08, ITF-2002-14,
[arXiv:hep-th/0204069].
\bibitem{Makeenko} Y.Makeenko, ITEP-TH-80-99, [arXiv:hep-th/0001047];\\
S.R. Das, Rev.Mod.Phys., {\bf 59} (1987) 235.
\bibitem{Parisi} G.Parisi,
Nucl.Phys., {\bf B100} (1975) 368.
\bibitem{Schnitzer} J.Schnitzer, Nucl. Phys., {\bf B109} (1976) 297.
\bibitem{reg} G.t'Hooft and M.J.G. Veltman, Nucl. Phys., {\bf B44} (1972) 189.
\bibitem{Arefeva} I.Ya.Aref'eva, Theor.Math.Phys., {\bf 29} (1976) 147; {\it ibid}
 {\bf 31} (1977) 3.
\bibitem{SchnitzerA} L.F.Abbott, J.S.Kang, H.J.Schnitzer, Phys. Rev., {\bf D13} (1976) 2212;\\
 T.Jaroszewicz, P.S.Kurzepa, Int.J.Mod.Phys., {\bf A8} (1993) 1613.
\bibitem{Anselmi} D.Anselmi, JHEP, {\bf 0006} (2000) 042,
[arXiv: hep-th/0005261].
\bibitem{Gracey1} J.A.Gracey, Int. J. Mod. Phys., {\bf A8} (1993) 2465-2486,
[arXiv:hep-th/9301123]; Phys. Lett., {\bf B317} (1993) 415-420,
[arXiv:hep-th/9309092].
\bibitem{Hooft} G.t'Hooft, Nucl. Phys., {\bf B61} (1973) 455.
\bibitem{powerbehav} K.Dienes, E.Dudas, and
T.Gherghetta, Phys.Lett., {\bf B436} (1998) 55 [arXiv:
hep-ph/9803466]; Nucl.Phys., {\bf B537} (1998) 47
[arXiv:hep-ph/9806292 ].
\bibitem{Gracey2} J.A.Gracey, Phys. Lett., {\bf B318} (1993) 177,
 [arXiv:hep-th/9310063]; Phys. Lett., {\bf B373} (1996) 178, [arXiv:hep-ph/9602214].
\bibitem{Hooft1} G.t'Hooft, Nucl. Phys., {\bf B72} (1974) 461.
\bibitem{Cut} R.E.Cutkosky, J.Math.Phys., {\bf 1} (1960) 429.
\bibitem{Kr} N.V.Krasnikov, Phys.Lett., {\bf B273} (1991) 246; JETP Lett., {\bf 51} (1990)
4.
\bibitem{gravity} F. Canfora, Phys.Rev., {\bf D74} (2006) 064020,
[arXiv:hep-th/0608203]; Nucl.Phys., {\bf B731} (2005) 389,
[arXiv:hep-th/0511017]; \\
N.E.Bjerrum-Bohr, Nucl.Phys., {\bf B684} (2004) 209,
[arXiv:hep-th/0310263]; \\
A.Strominger, Phys.Rev., {\bf D24} (1981) 3082;\\
E. Tomboulis, Phys.Lett., {\bf B70} (1977) 361.
\bibitem{Ward} B.F.L.Ward, arXiv:hep-ph/0607198; Mod.Phys.Lett., {\bf A17} (2002) 2371; ibid.{\bf A19} (2004) 143; J.Cos.Astropart.Phys., {\bf 0402} (2004) 011.
\bibitem{YFS} D.R.Yennie, S.C.Frautschi and H.Suura, Ann. Phys., {\bf 13}
(1961) 379; \\
K. T. Mahanthappa, Phys. Rev., {\bf 126} (1962) 329.
\end{thebibliography}
\end{document}